\documentclass[aps,twocolumn,pra]{revtex4-1}
\usepackage{graphics}
\usepackage{dcolumn}
\usepackage{bm}
\usepackage{amsmath}
\usepackage{hyperref}
\hypersetup{colorlinks=true, urlcolor=blue}
\usepackage{color}

\usepackage{newlfont}
\usepackage{amssymb}
\usepackage{amsfonts}
\usepackage{amsmath}
\usepackage{graphicx}
\usepackage{bm}

\def \be{\begin{equation}}
\def \ee{ \end{equation} }

\begin{document}

\definecolor{red}{rgb}{1,0,0}
\title{Do Large Number of Parties Enforce Monogamy in All Quantum Correlations?}

\author{Asutosh Kumar, R. Prabhu, Aditi Sen(De), and Ujjwal Sen}

\affiliation{Harish-Chandra Research Institute, Chhatnag Road, Jhunsi, Allahabad 211 019, India}

\begin{abstract}
Monogamy is a non-classical property that restricts the sharability of 
quantum correlation among the constituents of a multipartite quantum
system. Quantum correlations may satisfy or violate monogamy for 
 quantum states. Here we provide evidence that almost all pure quantum
states of systems consisting of a large number of subsystems are
monogamous with respect to all quantum correlation measures of both the
entanglement-separability and the information-theoretic paradigms,
indicating that the volume of the monogamous pure quantum states increases with
an  increasing number of parties. Nonetheless, we identify important
classes of pure states that remain non-monogamous with  respect to quantum discord
and quantum work-deficit, irrespective of the number of qubits. We
find conditions for which a given quantum  correlation measure
satisfies vis-\`a-vis violates monogamy.

\end{abstract}

\maketitle

\section{Introduction}
Correlations,  classical as well as quantum,  present in quantum systems play a significant role in quantum physics. 
Differentiating quantum correlations \cite{Horodecki09, ModiDiscord} from  classical ones have created a lot of attention, since it has been established that quantum ones can be a useful resource for quantum information protocols including those in quantum communication and possibly quantum computation. Moreover, it turns out to be an effective tool to detect cooperative quantum phenomena in many-body physics \cite{Lewenstein07, Amico08}. Current technological developments  ensure detection of  quantum correlations in several physical systems like photons \cite{XYZphoton}, ions \cite{XYZtrapion},  optical lattices \cite{Treutlein}, in the laboratories by using techniques like Bell inequalities \cite{inequality},  tomography \cite{tomography}, and  entanglement witnesses \cite{witness}.

In the case of bipartite systems, quantum correlations can  broadly be categorized into two groups: entanglement measures and information-theoretic ones. Entanglement  measures include, to name a few, entanglement of formation \cite{eof}, distillable entanglement \cite{eof,distillable}, logarithmic negativity \cite{VidalWerner}, relative entropy of entanglement \cite{VPRK}, etc., while quantum discord \cite{hv,oz}, quantum work-deficit \cite{workdeficit} are  information-theoretic ones. 


Monogamy forms a connecting theme in the exquisite variety in the space of quantum correlation measures. In simple terms, monogamy dictates that if two quantum systems are highly quantum correlated with respect to some quantum correlation measure, then they cannot be significantly quantum correlated with a third party with respect to that measure. It should be noted that monogamy is a non-classical concept and classical correlation can violate monogamy to the maximal extent. These qualitative statements have been quantified \cite{ekert,measent,CKW}, and while quantum correlations, in general, are expected to be qualitatively monogamous, they may violate it for some states, when quantitatively probed. In particular, while the square of the concurrence is monogamous for all multiqubit systems and the square of the negativity is monogamous for all three-qubit pure states \cite{CKW,negativitysquare,Osborne,monogamyN}, entanglement of formation, logarithmic negativity, quantum discord, and quantum work-deficit are known to violate monogamy already for three-qubit pure states \cite{amaderdiscord,giorgi,Fanchini,Dagmar,Salini}. 

Along with being a way to provide structure to the space of quantum correlation measures, the concept of monogamy is crucially important for security of quantum cryptography \cite{cryptoreview}. Moreover, they also provide a method for quantifying quantum correlations in multiparty scenarios, where it is usually difficult to conceptualize and compute quantum correlations \cite{ekert,measent,CKW,negativitysquare,Osborne,monogamyN,amaderdiscord,giorgi,Fanchini,Dagmar,Salini,MaheshExp}. 

In this paper,  we address the question of monogamy of bipartite quantum correlations for quantum states of an arbitrary number of parties. 
We prove a result to identify properties of bipartite quantum correlation measures
which are sufficient for that measure to follow the monogamy relation for all states. In particular, we show that if entanglement of formation is monogamous for a pure quantum state of an arbitrary number of parties, any bipartite ``good'' entanglement measure is also monogamous for that state.
Furthermore, we perform
numerical simulations 
with randomly chosen pure multiqubit quantum states over the uniform Haar measure, which clearly indicate that all the computable bipartite quantum correlation measures for almost all states become (quantitatively) monogamous with the increase in the number of parties.
Composing the analytical results with the numerical simulations, it follows that, for example, the relative entropy of entanglement and distillable entanglement, both non-computable but crucially important quantum information quantities, are (quantitatively) monogamous for almost all pure states of four or more qubits.  We also show that there are classes of multiparty pure quantum states that are non-monogamous for an arbitrary number of parties for certain quantum correlations. These classes, which have zero Haar volumes and hence are not covered in the random Haar searches, include the multiparty $W$ \cite{Wref},  the Dicke states \cite{dickestates}, and the symmetric states, and the corresponding quantum correlations are quantum discord and quantum work-deficit. We provide sufficient conditions for a multiparty quantum state to be non-monogamous. Precisely, we show that the multiqubit  states with vanishing tangle \cite{CKW} violate the monogamy relation for quantum discord with a certain nodal observer, provided the sum of the unmeasured conditional entropies of the nodal observer, conditioned on the non-nodal observers, is negative.



The rest of the paper is organized as follows. In Sec. \ref{sec:QCmeas}, we have given the definitions of the quantum correlation measures chosen for our study. 
In Sec. \ref{sec:monogamy}, we recapitulate the concept of monogamy for an arbitrary quantum correlation measure and numerically establish that all quantum correlation measures eventually become monogamous for almost all pure states with the increase in the number of parties. The zero Haar volume regions containing non-monogamous states are identified in Sec. \ref{sec:analytic} where we also find sufficient conditions for violation of monogamy for given multisite quantum states. We present a conclusion in Sec. \ref{sec:conclusion}. 


\section{Quantum Correlation Measures}
\label{sec:QCmeas}

Over the years, many quantum correlation measures have been proposed to quantify and characterize quantum correlation in quantum systems, consisting of two subsystems \cite{Horodecki09,ModiDiscord}.
The general properties that such bipartite measures should exhibit, have been extensively studied.
The sharing of bipartite quantum correlations among the subsystems of a multiparticle system plays an important role in this regard. In particular, it has been realized that it is important to understand the monogamy properties of these measures. We will take up this issue in the succeeding sections. In this section, we present brief definitions of the quantum correlation measures that we use in the succeeding sections.
Bipartite quantum correlation measures fall into two broad paradigms: (1) entanglement-separability measures and (2) information-theoretic quantum correlation measures. In this section, we define a few quantum correlation measures from both the paradigms.

\subsection{Measures of entanglement-separability paradigm}
\label{Sec:esmeas}

Here we consider the quantum correlation measures which vanish for separable states.
Moreover, they,  on average, do not increase under local quantum operations and classical communication (LOCC). Such quantum correlation measures belong to the entanglement-separability paradigm. We consider three  quantum correlation measures, viz., entanglement of formation, concurrence, and logarithmic negativity, within this  paradigm. 

\subsubsection{Entanglement of formation and concurrence}
\label{Sec:eof}

The entanglement of formation (EoF) \cite{concurrence} of a bipartite quantum state is the average number of singlets, \(\frac{1}{\sqrt{2}}(|01\rangle - |10\rangle)\),  that is required to prepare a single copy of the state by LOCC, assuming that the EoF of pure bipartite states are given by their local von Neumann entropies. Here \(|0\rangle\) and \(|1\rangle\) form an orthonormal qubit basis. For an arbitrary two-qubit state, \(\rho_{AB}\), there exists a closed form of the entanglement of formation \cite{concurrence}, in terms of the concurrence, ${\cal C}$, as 
\begin{equation}
\mathcal{E}(\rho_{AB})=h\left(\frac{1+\sqrt{1-\mathcal{C}^2(\rho_{AB})}}{2}\right),
\end{equation}
where $h(x)=-x\log_2x-(1-x)\log_2(1-x)$ is the Shannon (binary) entropy. Note that EoF is a concave function of $\mathcal{C}^2$, lies between 0 and 1, and vanishes for separable states.

Concurrence  can also be used to quantify entanglement for all two-qubit states \cite{concurrence}. 
For any two-qubit state, \(\rho_{AB}\), the concurrence is given by  
${\cal C}(\rho_{AB})=\mbox{max}\{0,\lambda_1-\lambda_2-\lambda_3-\lambda_4\}$, where the
$\lambda_i$'s are the square roots of the eigenvalues of $\rho_{AB}\tilde{\rho}_{AB}$ in decreasing order and 
$\tilde{\rho}_{AB}=(\sigma_y\otimes\sigma_y)\rho_{AB}^*(\sigma_y\otimes\sigma_y)$, with the complex conjugation being taken
in the computational  basis. $\sigma_y$ is the Pauli spin matrix.  For pure two-qubit states, $|\psi_{AB}\rangle$, the concurrence is given by $2\sqrt{\textrm{det} \rho_A}$, where $\rho_A$ is the subsystem density matrix obtained by tracing over the $B$-part from the two-qubit pure state $|\psi_{AB}\rangle$.
In case of pure states in $2\otimes d$, the concurrence is again given by $2\sqrt{\textrm{det} \rho_A}$ due to Schmidt decomposition. For mixed states in $2\otimes d$, one can use the convex roof extension to calculate the same.

\subsubsection{Logarithmic negativity}
\label{entanglement}

Logarithmic negativity (LN) \cite{VidalWerner} is another measure which belongs to the entanglement-separability paradigm. It is defined in terms of the negativity, \({\cal N}(\rho_{AB})\), of a bipartite state \(\rho_{AB}\). It is defined as the absolute value of the sum of the negative eigenvalues of \(\rho_{AB}^{T_{A}}\),
 where \(\rho_{AB}^{T_{A}}\) denotes the partial transpose of \(\rho_{AB}\) with respect to the \(A\)-part \cite{Peres_Horodecki}. It can be expressed as
 \begin{equation}
  {\cal N}(\rho_{AB})=\frac{\|\rho_{AB}^{T_A}\|_1-1}{2},
 \end{equation}
 where $\|M\|_1 \equiv \mbox{tr}\sqrt{M^\dag M}$ is the trace-norm of the matrix $M$.
The logarithmic negativity is defined as
\begin{equation}
E_{\cal N}(\rho_{AB}) = \log_2 [2 {\cal N}(\rho_{AB}) + 1].
\label{eq:LN}
\end{equation}
For two-qubit states, a strictly positive LN implies that the state is entangled and distillable \cite{Peres_Horodecki, Horodecki_distillable}, whereas 
a vanishing LN implies that the state is separable \cite{Peres_Horodecki}.

\subsection{Information-theoretic quantum correlation measures}

In this subsection, we will briefly describe two measures of quantum correlation chosen from the information-theoretic perspective. Although many of the quantum information protocols are assisted by quantum entanglement, there are several protocols for which presence of entanglement is not required \cite{nlwe,KnillLaflamme,Animesh,others,MLSM}. Information-theoretic quantum correlation measures may potentially explain such phenomena. 
Unlike entanglement-based quantum correlation measures, there does not exist closed forms for any information-theoretic measure. However, in many of the cases, it is possible to calculate them numerically.

\subsubsection{Quantum discord}
\label{discord}

Quantum discord for a bipartite state $\rho_{AB}$ is defined as the difference between the total correlation and the classical correlation of the state. The total correlation is defined as the quantum mutual information of  \(\rho_{AB}\), which is given by \cite{qmi} (see also \cite{Cerf, GROIS})
\begin{equation}
\label{qmi}
\mathcal{I}(\rho_{AB})= S(\rho_A)+ S(\rho_B)- S(\rho_{AB}),
\end{equation}
where $S(\varrho)= - \mbox{tr} (\varrho \log_2 \varrho)$ is the von Neumann entropy of the quantum state \(\varrho\). The classical correlation is based on the conditional entropy, and is defined as
\begin{equation}
\label{eq:classical}
 {\cal J}^{\leftarrow}(\rho_{AB}) = S(\rho_A) - S(\rho_{A|B}). 
\end{equation}
Here,
\begin{equation}
 S(\rho_{A|B}) = \min_{\{B_i\}} \sum_i p_i S(\rho_{A|i})
\end{equation}
is the conditional entropy of \(\rho_{AB}\), conditioned on a measurement performed by \(B\) with a rank-one projection-valued measurement \(\{B_i\}\),
producing the states  
\(\rho_{A|i} = \frac{1}{p_i} \mbox{tr}_B[(\mathbb{I}_A \otimes B_i) \rho (\mathbb{I}_A \otimes B_i)]\), 
with probability \(p_i = \mbox{tr}_{AB}[(\mathbb{I}_A \otimes B_i) \rho (\mathbb{I}_A \otimes B_i)]\).
\(\mathbb{I}\) is the identity operator on the Hilbert space of \(A\). Hence the discord can be calculated as
\cite{hv,oz}
\begin{equation}
\label{eq:discord}
{\cal D}^{\leftarrow}(\rho_{AB})= {\cal I}(\rho_{AB}) - {\cal J}^{\leftarrow}(\rho_{AB}).
\end{equation}
Here, the superscript ``$\leftarrow$" on ${\cal J}^{\leftarrow}(\rho_{AB})$ and ${\cal D}^{\leftarrow}(\rho_{AB})$ indicates that the measurement is performed on the subsystem $B$ of the state $\rho_{AB}$. Similarly, if measurement is performed on the subsystem $A$ of the state $\rho_{AB}$, one can define a quantum discord as
 \begin{equation}
\label{eq:discordA}
{\cal D}^{\rightarrow}(\rho_{AB})= {\cal I}(\rho_{AB}) - {\cal J}^{\rightarrow}(\rho_{AB}).
\end{equation}

\subsubsection{Quantum work-deficit}
\label{sec:workdeficit}

Quantum work-deficit \cite{workdeficit} is a quantum correlation measure also belonging to the information-theoretic paradigm. It is defined as the difference between the amount of pure states that can be extracted under global operations and pure product states that can be extracted under local operations, in closed systems for which addition of the corresponding pure states are not allowed.

 The number of pure qubits that can be extracted from $\rho_{AB}$ by ``closed global operations'' (CGO) is given by
\[I_G (\rho_{AB})= N - S(\rho_{AB}),\]
 where $N = \log_2 (\dim {\cal H})$. Here, CGO are any sequence of unitary operations and dephasing of the given state $\rho_{AB}$ by using a set of projectors $\{P_i\}$, i.e., $\rho \rightarrow \sum_i P_i \rho_{AB} P_i$,  
where $P_iP_j = \delta_{ij} P_i$, $\sum_i P_i = \mathbb{I}$, with $\mathbb{I}$ being the identity operator on the Hilbert space ${\cal H}$ on which $\rho_{AB}$ is defined. 

The number of qubits that can be extracted from a bipartite quantum state $\rho_{AB}$ under ``closed local operations and classical communication''(CLOCC), is given by 
\begin{equation}
I_L(\rho_{AB}) = N - \inf_{\Lambda \in CLOCC} [S(\rho{'}_A) + S(\rho{'}_B)],
\end{equation}
where $S(\rho{'}_A) = S(\mbox{tr}_B (\Lambda (\rho_{AB})))$  and $S(\rho{'}_B) = S(\mbox{tr}_A (\Lambda (\rho_{AB})))$. 
Here CLOCC can be local unitary, local dephasing, and sending dephased state from one party to another.

The quantum work-deficit is the difference between the work, $I_G (\rho_{AB})$,  extractable by CGO,  and that by CLOCC, $I_L (\rho_{AB})$:
\begin{equation}
 \Delta(\rho_{AB}) = I_G(\rho_{AB}) - I_L(\rho_{AB}).
\end{equation}
Since it is inefficient to compute this quantity for arbitrary states, we restrict our analysis only to CLOCC, where measurement is done at any one of the subsystems. One can show that one-way work-deficit is the same as quantum discord for bipartite states with maximally mixed marginals.

\section{Status of Monogamy of Quantum Correlations for Arbitrary Number of Parties}
\label{sec:monogamy}

In this section, we begin by formally introducing the concept of monogamy and the relations that a quantum correlation measure must satisfy, for it to be monogamous for a given quantum state of an arbitrary number of parties. We then 
try to find the extent to which bipartite quantum correlation measures satisfy the (quantitative) monogamy relation.

\subsection{Monogamy of quantum correlations}
\label{MonoQC}

Let ${\cal Q}$ be a bipartite quantum correlation measure. An \(n\)-party quantum state, $\rho_{12\cdots n}$, is said to be (quantitatively) monogamous under the quantum correlation measure ${\cal Q}$, if  it follows the inequality, given by \cite{CKW}
\begin{equation}
 {\cal Q}(\rho_{12})+{\cal Q}(\rho_{13})+\cdots+{\cal Q}(\rho_{1n})\leq{\cal Q}(\rho_{1(2\cdots n)}).
\end{equation}
Otherwise, it is non-monogamous. Here, $\rho_{12}=\mbox{tr}_{3\ldots n}(\rho_{12\cdots n})$, etc. and ${\cal Q}(\rho_{1(2\cdots n)})$ denotes the quantum correlation ${\cal Q}$ of $\rho_{12\cdots n}$ in the $1:2\cdots n$ bipartition. We will also denote ${\cal Q}(\rho_{12})$, etc. as ${\cal Q}_{12}$, etc. and 
${\cal Q}(\rho_{1(2\cdots n)})$ as ${\cal Q}_{1(2\cdots n)}$. The party ``\(1\)" can be referred to as the ``nodal" observer. In this respect,  one can define the ``\({\cal Q}\)-monogamy score" \cite{manab} for the \(n\)-party state, $\rho_{12\cdots n}$, as
\begin{equation}
\label{eq:monoscore}
\delta_{{\cal Q}} = {\cal Q}_{1(2\cdots n)} - \sum _{i=2}^{n} {\cal Q}_{1i}. 
\end{equation}
Non-negativity of \(\delta_{{\cal Q}}\) for all quantum states implies monogamy of \({\cal Q}\). For instance, the square of the concurrence has been shown to be monogamous \cite{CKW, Osborne} for all quantum states for an arbitrary number of qubits.  However, there exist other measures like entanglement of formation, quantum discord, and quantum work-deficit which are known to be non-monogamous, examined primarily for pure three-qubit states \cite{measent,amaderdiscord, giorgi, Salini}. See also \cite{Fanchini,Dagmar,GMS,RajagopalRendell,usha,FanchiniEof,Sandersrev, mono13}.


\subsection{Appearance of monogamy as generic in large systems}
\label{subsec:monoarb}

\begin{figure}%
\begin{center}
\includegraphics[angle=0,width=1\columnwidth]{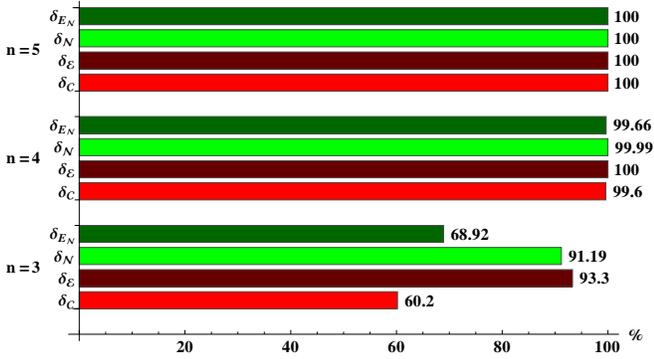}
\caption{(Color online) Percentage bar-diagram for monogamy scores of entanglement measures for pure $n$-party states. $10^5$ random pure quantum states are generated for each $n$. All notations are the same as for Table \ref{table:ent-mono-percent}.}
\label{fig-HistoPlot_EntMeas}
\end{center}
\end{figure}

In most of the previous works on monogamy, the status of the monogamy relation for different quantum correlation measures has been considered only for three-qubit pure states. 
Exceptions include the square of the concurrence, which is proven to be monogamous for any number of qubits \cite{Osborne}. Here we address the question of monogamy for an arbitrary number of parties for an arbitrary bipartite quantum correlation measure. Before presenting the general results, let us first consider the numerical data that we obtain for different bipartite measures, which strongly indicate that the results for tripartite systems can not be generalized to systems with a large number of parties. In particular, numerical simulations indicate that \emph{all} quantum correlation measures become monogamous for almost all multiparty pure quantum  states of a relatively moderate number of parties. Monogamous quantum states are known to be useful in several applications, including quantum cryptography. The relatively moderate number of parties -- five 
-- ensure that the results are relevant to the experimental techniques currently available, both with photons as well as with massive particles \cite{XYZphoton,XYZtrapion,Treutlein}. The 
status of the monogamy relation, as obtained via the numerical simulations, for all computable entanglement measures are summarized below in Table \ref{table:ent-mono-percent}. It is clear from  Table \ref{table:ent-mono-percent} that several entanglement measures which are  non-monogamous for three-qubit pure states, become monogamous, when one increases the number of parties by a relatively moderate amount. Some of the results from Table \ref{table:ent-mono-percent} are also depicted in 
Fig. \ref{fig-HistoPlot_EntMeas}. 

\begin{table}[ht]
\centering
\begin{tabular}{cccccccc}
\hline 
$n$ & $\delta_{\cal C}$ & $\delta_{\cal E}$ & $\delta_{{\cal E}^2}$ & $\delta_{\cal N}$ & $\delta_{{\cal N}^2}$ & $\delta_{E_{\cal N}}$ & $\delta_{E^2_{\cal N}}$ \\[0.5ex]
\hline
3 & 60.2 & 93.3 & 100 & 91.186 & 100 & 68.916 & 100\\[0.5ex]
4 & 99.6 & 100 & 100 & 99.995 & 100 & 99.665 & 100\\[0.5ex]
5 & 100 & 100 & 100 & 100 & 100 & 100 & 100\\[1ex]
\hline
\end{tabular}
\caption{Monogamy percentage table for entanglement measures. We randomly choose $10^5$  pure quantum states uniformly 
according to the Haar measure over the complex $2^n$-dimensional Hilbert space for each $n$, for $n=3,4,\ \textrm{and}\ 5$. Here $n$, therefore, denotes the number of qubits forming the system from which the pure quantum states are chosen. $\delta_{\cal C}, \delta_{\cal E}, \delta_{\cal N}, \delta_{E_{\cal N}}$ respectively denote the monogamy scores of concurrence, EoF, negativity and logarithmic negativity, while $\delta_{{\cal E}^2}, \delta_{{\cal N}^2}, \delta_{E_{\cal N}^2}$  are the monogamy scores of the squares of these measures. The numbers shown are percentages of the randomly chosen states that are monogamous for that case.
}\label{table:ent-mono-percent}
\end{table}

Before moving to the other quantum correlation measures, let us prove here a sufficient condition that has to be satisfied by any entanglement measure for it to be monogamous. 

\noindent\textbf{Theorem 1:(Monogamy for given states.)} \textit{If entanglement of formation is monogamous for a pure quantum state of an arbitrary number of parties, any bipartite ``good" entanglement measure is also monogamous for that state.}\\
\textbf{Remark.} In Ref. \cite{horodecki-limit}, bipartite entanglement measures satisfying certain reasonable axioms, were referred to as ``good" measures, and were shown to be bounded above by the entanglement of formation and equal to the local von Neumann entropy for pure bipartite states \cite{measent}. Here, we slightly broaden the definition, and call an entanglement measure as ``good" if it is lower than or equal to the entanglement of formation and it is equal to the local von Neumann entropy for pure bipartite states.

\noindent\texttt{Proof.} Let $|\psi_{12\cdots n}\rangle$ be a multipartite quantum state consisting of $n$ parties and $\delta_{\cal E}$, the monogamy score of entanglement of formation, be non-negative for the $n$-party state. Consider now any ``good" bipartite entanglement measure ${\cal Q}$.
Therefore, when entanglement of formation is monogamous, we have
\begin{equation}
\mathcal{Q}(|\psi_{1(2\cdots n)}\rangle)= S(\rho^{\psi}_1)= \mathcal{E}(|\psi_{1(2\cdots n)}\rangle)\geq \sum_{j=2}^n\mathcal{E}_{1j} \geq \sum_{j=2}^n\mathcal{Q}_{1j}
 \label{theorem}
 \end{equation}
Hence the proof. (Here $\rho^{\psi}_1=\mbox{tr}_{2\cdots n}|\psi\rangle\langle \psi|$.)
\hfill $\blacksquare$

Table \ref{table:ent-mono-percent} shows that entanglement of formation is monogamous for almost all pure states of four qubits. Utilizing this result along with that from Theorem 1, we obtain that relative entropy of entanglement, regularized relative entropy of entanglement \cite{regREE}, entanglement cost \cite{eof,Entcost}, distillable entanglement, all of which are not generally computable, are monogamous for almost all pure states of four or more qubits.

\begin{figure}%
\begin{center}
\includegraphics[angle=0,width=1\columnwidth]{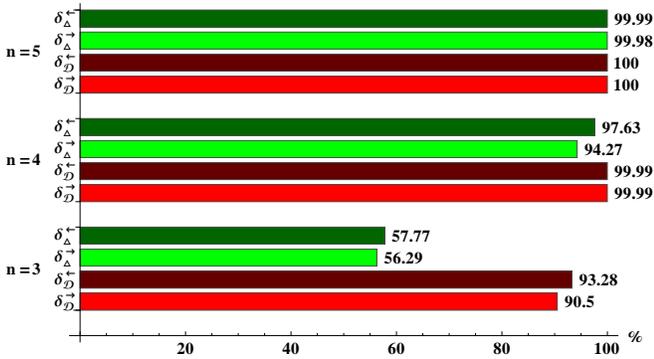}
\caption{(Color online) Percentage bar-diagram for monogamy scores of information-theoretic quantum correlation measures for pure $n$-party states. $10^5$ random pure quantum states are generated for each $n$. All notations are the same as in Table \ref{table:info-mono-percent}.}
\label{fig-HistoPlot_DisWDscore}
\end{center}
\end{figure}

Just like for entanglement measures, and as displayed in Table \ref{table:info-mono-percent}, percentages of randomly chosen pure states satisfying monogamy, increase also for all information-theoretic quantum correlation measures with the increase in the number of parties (see also  Fig. \ref{fig-HistoPlot_DisWDscore}). Note here that the square of quantum discord (precisely ${{\cal D}^{\leftarrow}}^2$) was shown to be monogamous \cite{discordsquare} 
for three-qubit pure states. For ease of notation, we often denote $\delta_{{{\cal D}^{\rightarrow}}^2}$,  $\delta_{\vartriangle^{\leftarrow}}$, etc., as $\delta_{{\cal D}^2}^{\rightarrow}$,  $\delta_{\vartriangle}^{\leftarrow}$, etc., respectively.  It is to be noted here that uniform Haar searches may tend to become inefficient when we consider a large number of parties  (cf. \cite{AWinter}). However, they are efficient for the few qubit systems that we consider here, especially for $n=3$ and $n=4$.


\begin{table}[ht]
\centering
\begin{tabular}{ccccccccc}
\hline 
$n$ & $\delta_{\cal D}^{\rightarrow}$ & $\delta_{{\cal D}^2}^{\rightarrow}$ & $\delta_{\cal D}^{\leftarrow}$ & $\delta_{{\cal D}^2}^{\leftarrow}$ 
& $\delta_{\vartriangle}^{\rightarrow}$  & $\delta_{{\vartriangle}^2}^{\rightarrow}$ & $\delta_{\vartriangle}^{\leftarrow}$ & $\delta_{{\vartriangle}^2}^{\leftarrow}$\\[0.5ex]
\hline 
3 & 90.5 & 100 & 93.28 & 100 & 56.29 & 88.10  & 57.77 & 89.56\\[0.5ex]
4 & 99.997 & 100 & 99.99 & 100 & 94.27 & 99.99 & 97.63 & 100\\[0.5ex]
5 & 100 & 100 & 100 & 100 & 99.98 & 100 & 99.99 & 100 \\[1ex]
\hline
\end{tabular}
\caption{Percentage table for quantum states satisfying the monogamy relation for information-theoretic paradigm measures. We randomly chose $10^5$ pure quantum states uniformly according to the Haar measure over the complex $2^n$-dimensional Hilbert space. 
The numbers shown are percentages of the randomly chosen states that are monogamous for that case.
}\label{table:info-mono-percent}
\end{table}

Let us now specify certain sets of properties that are sufficient for any quantum correlation measure to satisfy the monogamy relation for an arbitrary number of parties in arbitrary dimensions, when it is monogamous for smaller number of parties \cite{Osborne}. 
Let us consider an \(n\)-party state,  \( |\psi_{12\ldots n}\rangle\) in \(d\) dimensions, in which we  make the partition, in such a way that the final state is always  tripartite. In this case, we have the following theorem. \\

\noindent\textbf{Theorem 2: (Monogamy for given measures.)} \textit{If \({\cal Q}\) is monogamous for all tripartite quantum states in \(d \otimes d \otimes d_C \) where \(d_C =d^m,\ m\leq n-2\), then \({\cal Q}\) is monogamous for pure quantum states of the $n$ parties. 
}
 
\noindent\texttt{Proof.} 
Suppose that the dimension of the third party is \(d^{n-2}\), consisting of \((n-2)\) parties, say \(3, \ldots, n\), each being of dimension \(d\). The monogamy of such a \(3\)-party state, \(|\psi_{1:2:3 \ldots n}\rangle\), implies that
\begin{eqnarray}
\label{eq:recursion}
{\cal Q}(|\psi_{1(23\ldots n)}\rangle)   &\geq& {\cal Q}(\rho^{\psi}_{12}) + {\cal Q}(\rho^{\psi}_{1(3\ldots n)}) \nonumber \\
&\geq &  {\cal Q}(\rho^{\psi}_{12}) +   {\cal Q}(\rho^{\psi}_{13}) + {\cal Q}(\rho^{\psi}_{1(4\ldots n)}) \nonumber \\ 
&\geq &  {\cal Q}(\rho^{\psi}_{12}) +   {\cal Q}(\rho^{\psi}_{13}) + \sum_{k=4}^{n}{\cal Q}(\rho^{\psi}_{1k}),  
\end{eqnarray}
where the second inequality is obtained by applying the monogamy relation for the  tripartite state \(\rho^{\psi}_{1:3:4\ldots n}\), with the third party having \((n-3)\) parties, and the third inequality is by applying such monogamy relations recursively. Here, the local density matrices are denoted by $\rho^{\psi}$ with the appropriate suffixes determined by the parties that are not traced out.
\hfill $\blacksquare$


\noindent\textbf{Theorem 3:} \textit{If \({\cal Q}\) is monogamous for all tripartite pure quantum states in \(d \otimes d \otimes d_C \) where \(d_C =d^m,\ m\leq n-2\), then \({\cal Q}\) is monogamous for all quantum states, pure or mixed,  of the $n$ parties, provided \({\cal Q}\) is convex, and \({\cal Q}\) for mixed states is defined through the convex roof approach. 
}
 
\noindent\texttt{Proof.} 
Consider a mixed state $\rho_{123\ldots n}$ in the tripartition $1:2:3\ldots n$ and let $\{p_i,\, |\psi^i_{1(2\ldots n)}\rangle\}$ be the optimal decomposition that attains the convex roof of ${\cal Q}(\rho_{1(2\ldots n)})$.
Therefore,
\begin{eqnarray}
{\cal Q}(\rho_{1(23\ldots n)}) &=& {\cal Q}\left(\sum_i p_i|\psi^i_{1(2\ldots n)}\rangle\langle\psi^i_{1(2\ldots n)}|\right) \nonumber \\
&=& \sum_ip_i{\cal Q}(|\psi^i_{1(2\ldots n)}\rangle).
\end{eqnarray}
Due to the assumed monogamy over pure states, we have
\begin{eqnarray}
{\cal Q}(\rho_{1(2\ldots n)}) &\geq & \sum_i p_i \left({\cal Q}(\rho^{\psi^i}_{12})+{\cal Q}(\rho^{\psi^i}_{1(3\ldots n)})\right) \nonumber \\
&=&\sum_i p_i {\cal Q}(\rho^{\psi^i}_{12}) + \sum_i p_i  {\cal Q}(\rho^{\psi^i}_{1(3\ldots n)}),
\end{eqnarray}
which, due to convexity of ${\cal Q}$, reduces to 
\begin{eqnarray}
{\cal Q}(\rho_{1(2\ldots n)}) &\geq & {\cal Q}(\rho_{12}) +  {\cal Q}(\rho_{1(3\ldots n)}).
\end{eqnarray}
The result follows now by applying Theorem 2 and concavity of ${\cal Q}$. 
\hfill $\blacksquare$

\section{A zero measure class of non-monogamous states}
\label{sec:analytic}

In the preceding section, we presented evidence that almost all multiparty states for even a moderate number of parties are monogamous with respect to all quantum correlation measures. The qualification ``almost'' is important and necessary, firstly because the uniform Haar searches do not take into account of violations of the corresponding property (monogamy, here) on hypersurfaces (more generally, on sets of zero Haar measure). Secondly, and more constructively, we identify 
a class of multiparty states that are non-monogamous with respect to information-theoretic quantum correlation measures for an arbitrary number of parties. We begin by deriving an analytic relation, which will subsequently help us to identify the class of states.

For an arbitrary tripartite quantum state $\rho_{123}$, we have the relation \cite{Koashi-Winter}
\begin{equation}
{\cal S}(\rho_{1|3}) + D^{\leftarrow}(\rho_{13}) \geq \mathcal{E}(\rho_{12}),
\end{equation}
where ${\cal S}(\rho_{1|3}) = S(\rho_{13})-S(\rho_3)$ is the ``unmeasured conditional entropy" of $\rho_{13}$ conditioned on the party 3, and $\rho_{13},\, \rho_{12}$, and $\rho_{3}$ are local density matrices of $\rho_{123}$ of the corresponding parties.
Let us now consider an $n$-party  pure state $|\psi_{12\cdots n}\rangle$. Applying the above relation for an arbitrary tripartite partition of $|\psi_{12\cdots n}\rangle$,  
we obtain
\begin{equation}
{\cal S}(\rho^{\psi}_{1|j}) + D^{\leftarrow}(\rho^{\psi}_{1j}) \geq \mathcal{E}(\rho^{\psi}_{1i}),
\end{equation}
with $i\ne j$, and where the superscript $\psi$ on the local density matrices indicates that they are obtained by tracing out the requisite parties from the state $|\psi_{12\ldots n}\rangle$. If we choose $i=j+1$ for all $n$ except  for $j=n$ and choose $i=2$ for $j=n$, we have
\begin{equation}
 \sum_{j=2}^n\mathcal{E}(\rho^{\psi}_{1j}) \leq \sum_{j=2}^n\left(\mathcal{S}(\rho^{\psi}_{1|j})+{\cal D}^{\leftarrow}(\rho^{\psi}_{1j})\right).
 \label{eq:eofdisc}
\end{equation}

We now move to prove a theorem by using the above inequality. The tangle for a multiqubit quantum state $\rho_{12\ldots n}$ is denoted by $\tau(\rho_{12\ldots n})$ and is defined as $ \mathcal{C}^2(\rho_{1(2\cdots n)})-\sum_{j=2}^n\mathcal{C}^2(\rho_{1j})$.

\noindent\textbf{Theorem 4:} {\em Multiparty pure states with vanishing tangle violate the monogamy relation for quantum discord with a certain nodal observer, provided the sum of the unmeasured  conditional entropies, conditioned on all non-nodal observers, is negative. }

\noindent\texttt{Proof.} Let us consider an $n$-party  state $|\psi_{12\cdots n}\rangle$ for which the tangle vanishes, i.e., the state saturates the monogamy relation for  ${\cal C}^2$. Hence
$\sum_{j=2}^n\mathcal{C}^2(\rho^{\psi}_{1j})= \mathcal{C}^2(|\psi_{1(2\cdots n)}\rangle)$,
where by
 we have 
\begin{equation}
 \sum_{j=2}^n\mathcal{E}(\rho^{\psi}_{1j}) \geq  \mathcal{E}(|\psi_{1(2\cdots n)}\rangle).
\end{equation}
Since $\mathcal{E}(|\psi_{1(2\cdots n)}\rangle)= S(\rho^{\psi}_1)=\mathcal{D}^{\leftarrow}(|\psi_{1(2\cdots n)}\rangle)$, by using the  inequality in  (\ref{eq:eofdisc}),
we have 
\begin{eqnarray}
 &&\sum_{j=2}^n\left({\cal S}(\rho^{\psi}_{1|j})+{\cal D}^{\leftarrow}(\rho^{\psi}_{1j})\right) \geq \sum_{j=2}^n\mathcal{E}(\rho^{\psi}_{1j}) \nonumber\\
 &&\geq \mathcal{E}(|\psi_{1(2\cdots n)}\rangle)=  S(\rho^{\psi}_1)=\mathcal{D}^{\leftarrow}(|\psi_{1(2\cdots n)}\rangle).
\end{eqnarray}
Therefore, the discord monogamy score has the following bound:
\begin{equation}
 \delta_{\cal D}^{\leftarrow}=\mathcal{D}^{\leftarrow}(|\psi_{1(2\cdots n)}\rangle)-\sum_{j=2}^n{\cal D}^{\leftarrow}(\rho^{\psi}_{1j})
 \leq \sum_{j=2}^n {\cal S}(\rho^{\psi}_{1|j}).
 \label{eq:discordpolygamy}
\end{equation}
Hence, if states with  vanishing tangle, additionally satisfies $\sum_{j=2}^n\mathcal{S}(\rho^{\psi}_{1|j}) <0$, quantum discord is non-monogamous for those states. \hfill $\blacksquare$

The non-monogamy of discord for three-qubit $W$ states \cite{amaderdiscord,giorgi} is a special case of Theorem 4.
It can be easily checked that the $n$-qubit W state, $|W_n\rangle$  \cite{Wref}, given by 
\begin{equation}
|W_n\rangle = \frac{1}{\sqrt{n}}(|0\ldots1\rangle + \ldots + |1\ldots0\rangle),
\end{equation}
remain non-monogamous with respect to quantum discord for an  arbitrary number of parties.

\begin{figure}%
\begin{center}
\includegraphics[angle=0,width=0.8\columnwidth]{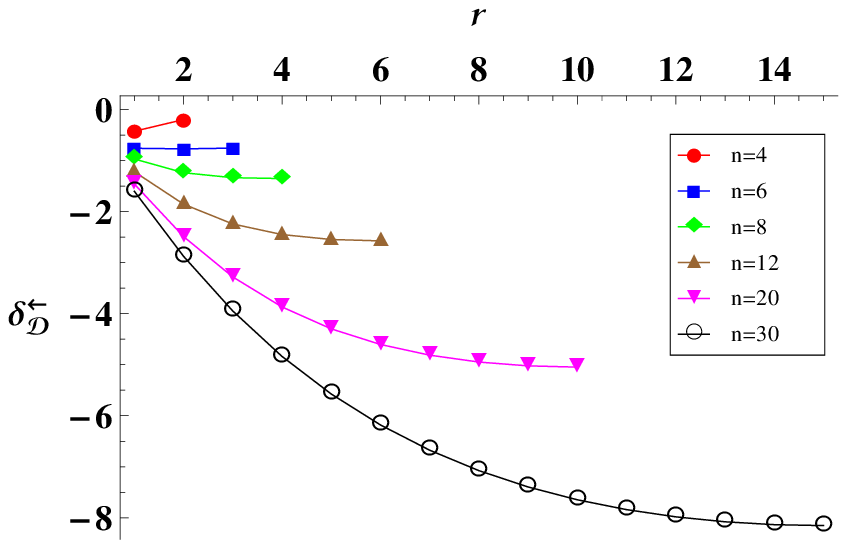}
\includegraphics[angle=0,width=0.8\columnwidth]{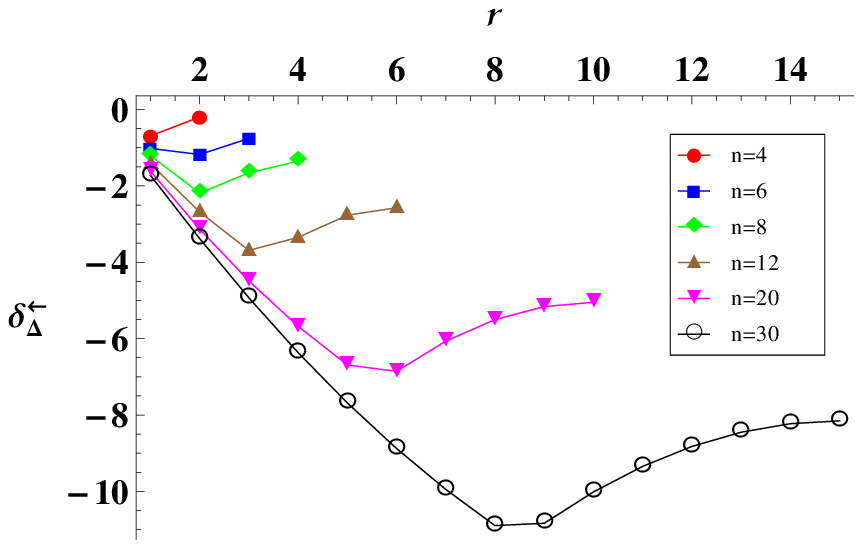}
\caption{(Color online) Top panel: Non-monogamy of Dicke states
with respect to quantum discord. For fixed $n\ (>6)$, $\delta_{\cal D}^{\leftarrow}$ decreases with increasing $r$ and  it decays exponentially for large $n$. Bottom panel: Non-monogamy of Dicke states with respect to quantum work-deficit. For fixed $n\ (>6)$, the trajectory of $\delta_{\vartriangle}^{\leftarrow}$ resembles an anharmonic potential well. It decays with increasing $r$ (upto $r \leq \left[\frac{n}{4}\right]$) and rises in the interval $\left[\frac{n}{4}\right]< r \leq \left[\frac{n}{2}\right]$. In both the panels, the horizontal axes are in terms of the number of excitations in the corresponding Dicke state, while the vertical axis in the top panel is in bits and that in the bottom panel is in qubits.}
\label{fig-JMstate_DisSco}
\end{center}
\end{figure}

\begin{figure}%
\begin{center}
\includegraphics[angle=0,width=0.8\columnwidth]{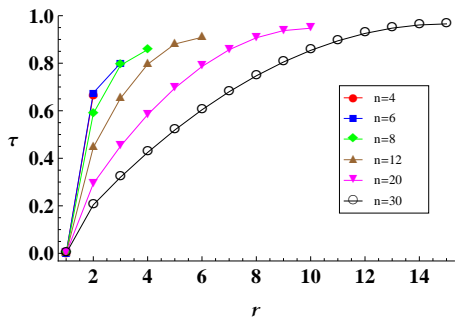}
\caption{(Color online) Tangle $(\tau)$
for Dicke states 
is plotted against the number of excitations, $r$ present in the $n$-party system. 
The tangle vanishes for all Dicke states with $r=1$ (i.e., for the $W$ states) and remains positive for Dicke states with $r>1$. Tangle is measured in ebits.}
\label{fig-JMstate_EntMeas}
\end{center}
\end{figure}

Let us now discuss some further classes of states which remain non-monogamous for quantum discord for arbitrary number of parties. 
Towards that end, consider the Dicke state \cite{dickestates}
\begin{equation}
 |W_{n}^r\rangle=\frac{1}{\sqrt{\binom{n}{r}}} \sum_{permuts}|\underbrace{00...0}_{n-r}\underbrace{11...1}_{r}\rangle,
 \label{eq:Dicke}
\end{equation}
where $\sum_{permuts}$ represents the unnormalized equal superposition  over all $\binom{n}{r}$ combinations of $(n-r)$ $|0\rangle$'s and $r$ $|1\rangle$'s.
We now examine the discord score of the above state.
Using the property that the Dicke state is permutationally invariant with respect to the subsystems, the optimization involved in computing quantum discord of the two-qubit reduced density matrices can be obtained analytically \cite{Chen}. Hence, an analytic expression of discord score for the Dicke states 
can be obtained and is given by
\begin{equation}
 \delta_{\cal D}^{\leftarrow}(|W_{n}^r\rangle)=S_1-(n-1)\Big(S_2- S_{12}+H(\{\lambda_{\pm}\})\Big),
\end{equation}
where
\begin{eqnarray}
  S_1&=&-\frac{r}{n}\log_2 \frac{r}{n}-(1-\frac{r}{n})\log_2 (1-\frac{r}{n}), \nonumber \\
  S_2&=&-(a+b)\log_2 (a+b)-(b+c)\log_2 (b+c), \nonumber \\
 S_{12}&=&-a\log_2 a-2b\log_2 2b-c\log_2 c, \nonumber \\
 \lambda_{\pm}&=&(1 \pm \sqrt{1-4(ab+bc+ca)})/2,
\end{eqnarray}
with $a=\frac{(n-r)(n-r-1)}{n(n-1)}$, $b=\frac{r(n-r)}{n(n-1)}$ and $c=\frac{r(r-1)}{n(n-1)}$. Note here that the tangle vanishes for the Dicke states for $r=1$. However it is non-vanishing for $r\neq 1$ and hence the previous theorem cannot be applied for the Dicke states with $r\neq 1$. The quantum discord and work-deficit scores of the Dicke states for various choices of $n$ with respect to excitations, $r$, are plotted in Fig. \ref{fig-JMstate_DisSco}. For comparison, the tangle of the Dicke states against $r$ for different $n$ is plotted in Fig. \ref{fig-JMstate_EntMeas}.

Although 
the Dicke states, for arbitrary $r$ and $n$, are non-monogamous with respect to discord and work-deficit,
the generalized Dicke states given by
\begin{equation}
 |W_{n}^r(\alpha_i^r)\rangle= \sum_i \alpha^r_i |\underbrace{00...0}_{n-r}\underbrace{11...1}_r\rangle
 \label{eq:genDicke}
\end{equation}
with the normalization $\sum_i |\alpha^r_i|^2=1$,  for $r>1$,
are largely monogamous. In Table \ref{table:dicke-mono-percent}, we list the percentage of randomly chosen states with positive quantum discord score and quantum work-deficit score of the generalized Dicke states as given in Eq. (\ref{eq:genDicke}) for $n=3,4,5,6$ qubits for some excitations, $r$. 

\begin{table}[ht]
\centering
\begin{tabular}{c c c c c c}
\hline
$n$ & 3  & 4  & 5  & 6 & 6 \\[0.5ex]
$r$ & 1 & 2  & 2  & 2 & 3 \\[0.5ex]
\hline \\[0.5ex]
$\delta_{\cal D}^{\rightarrow}$ & 0  & 94.86 & 99.29 & 99.80 & 100 \\[1ex]
$\delta_{\vartriangle}^{\rightarrow}$ & 27.19  & 66.77  & 71.46  & 72.01 & 88.25 \\[1ex]
\hline
\end{tabular}
\caption{Percentage table for the generalized Dicke states (see Eq. (\ref{eq:genDicke})) that satisfies monogamy for quantum discord and quantum work-deficit for $10^5$ randomly chosen pure states in each of the cases, according to the Haar measure over the corresponding space.}
\label{table:dicke-mono-percent} 
\end{table} 




As a final example, we now consider a general $n$-qubit symmetric state, given by
$|\psi_{GS}\rangle=\sum_{r=0}^na_r |W_n^r\rangle$.
We generate $10^5$ states randomly in the space of $n$-qubit symmetric states for $n=3,4,5,6$ uniformly according to the Haar measure. The monogamy scores for quantum discord and quantum work-deficit are calculated. The percentage of states which turned out to be monogamous in the different cases are depicted in Table \ref{table:gensymmet-mono-percent}. 
\begin{table}[ht]
\centering
\begin{tabular}{c c c c c}
\hline
$n$ & 3 & 4 & 5 & 6 \\[0.5ex]
\hline \\[0.5ex]
$\delta_{\cal D}^{\rightarrow}$ & 97.47 & 98.37 & 86.12 & 49.71 \\[1ex]
$\delta_{\cal D}^{\leftarrow}$ & 97.15 & 97.69 & 86.77 & 64.35 \\[1ex]
\hline \\[0.5ex]
$\delta_{\vartriangle}^{\rightarrow}$ & 81.40 & 81.49 & 56.41 & 26.40 \\[1ex]
$\delta_{\vartriangle}^{\leftarrow}$ & 78.97 & 77.77 & 61.19 & 41.48 \\[1ex]
\hline
\end{tabular}
\caption{Percentage table for the $n$-qubit symmetric states that satisfies monogamy for quantum discord and quantum work-deficit for $10^5$ randomly chosen pure states, for $n=3,4,5,6$.
}
\label{table:gensymmet-mono-percent} 
\end{table} 
Comparing now with the Tables \ref{table:ent-mono-percent} and \ref{table:info-mono-percent}, where general quantum states (not necessarily symmetric) in the same multiqubit spaces were considered, we find that in drastic contrast to those cases, the frequency of states which satisfies monogamy actually decreases with increasing number of qubits for the symmetric case. We have performed a finite-size scaling analysis, by assuming that all symmetric states will be monogamous for sufficiently large $n$. We find  log-log scaling, with the scaling law being 
$$p_n \approx p_c + n^{-\alpha},$$ 
where $p_n$ is the percentage of symmetric $n$-qubit states that are monogamous with respect to a given measure, ${\mathcal Q}$, $p_c \equiv p_{n \rightarrow \infty}$ (being assumed to be vanishing),  and $\alpha$ is the critical exponent of the scaling law. Based on the percentages obtained in the Haar searches, we calculated the critical exponents, and are displayed them in Table \ref{table:gensymmet-mono-critiexponent}.
\begin{table}[ht]
\centering
\begin{tabular}{cc}
\hline
${\mathcal Q}$ & $\alpha$ \\
\hline
$\delta_{\cal D}^{\rightarrow}$ & 0.8715 \\
$\delta_{\cal D}^{\leftarrow}$ & 0.5523\\
\hline
$\delta_{\vartriangle}^{\rightarrow}$ & 1.5351 \\
$\delta_{\vartriangle}^{\leftarrow}$ & 0.8960\\
\hline
\end{tabular}
\caption{The critical exponents of the scaling law of monogamous states among the  symmetric states, for quantum discord and quantum work-deficit.
}
\label{table:gensymmet-mono-critiexponent} 
\end{table} 
It should be noted here that all the classes of pure states considered in this section fall in a set of zero Haar measure in the space of all pure quantum states, for a given $n$-qubit space. This is also true for all symmetric pure states. It is plausible that symmetric mixed states form a non-zero, perhaps fast-decaying, volume of monogamous multiparty quantum states within the space of all quantum states, for large systems.


\section{conclusion}
\label{sec:conclusion}

%

In quantum communication protocols, in particular in quantum key sharing, it is desirable to detect and control external noise like eavesdropping. In this case, the concept of monogamy comes as a savior, as it does not allow an arbitrary sharing of quantum correlation among subsystems of a larger system. Thus identifying and quantifying which quantum correlation measures are monogamous for the given states, and under what conditions, become extremely important.
It is well-known that bipartite quantum correlation measures are, in general, not monogamous for arbitrary tripartite pure states.
We have shown that a quantum correlation measure which is 
non-monogamous for a substantial section of tripartite quantum states, becomes monogamous for almost all quantum states of $n$-party systems, with $n$ being only slightly higher than 3. We have also identified sets of zero Haar measure in the space of all multiparty quantum states that remain non-monogamous for an arbitrary number of parties. Apart from providing an understanding on the structure of space of quantum correlation measures, and their relation to the underlying space of multiparty quantum states, our results may shed more light on the methods for choosing quantum systems for secure quantum information protocols, especially in large quantum networks.


\begin{acknowledgments}
We acknowledge discussions with Arul Lakshminarayan, M. S. Ramkarthik, and Andreas Winter.
RP acknowledges an INSPIRE-faculty position at the Harish-Chandra Research Institute (HRI) from the Department of Science and Technology, Government of India.
We acknowledge computations performed at the cluster computing facility at HRI. 
\end{acknowledgments}

\end{document}